\documentclass[preprint,amsmath,amssymb,showkeywords,showpacs,groupedaddress]{revtex4}
\usepackage[pdftex]{color}
\usepackage{graphicx}
\usepackage{dcolumn}
\usepackage{bm}
\usepackage{amsmath}
\usepackage[letterpaper,left=20mm,right=20mm,top=30mm,bottom=20mm]{geometry}
\usepackage{textcomp}
\usepackage[pdftex]{color}

\begin{document}
\draft
\title{Quantum phase transition in easy-axis antiferromagnetic Heisenberg spin-1 chain}
\date{\today}
\author{Jie Ren}
\author{Shiqun Zhu\footnote{Corresponding author,E-mail: szhu@suda.edu.cn}}

\affiliation{School of Physical Science and Technology, Suzhou
University, Suzhou, Jiangsu 215006, People's Republic of China}

\begin{abstract}

The fidelity and entropy in an easy-axis antiferromagnetic
Heisenberg spin-1 chain are studied numerically. By using the method
of density-matrix renormalization-group, the effects of anisotropy
on fidelity and entanglement entropy are investigated. Their
relations with quantum phase transition are analyzed. It is found
that the quantum phase transition from Haldane spin liquid to
N\'{e}el spin solid can be well characterized by the fidelity. The
phase transition can be hardly detected by the entropy. But it can
be successfully detected by the first deviation of the entropy.
\end{abstract}

\pacs{03.67.-a, 03.65.Ud, 75.10.Pq}

\maketitle

\section{Introduction}

In condense matter physics, quantum phase transitions imply
fluctuations, which happened at the zero temperature~\cite{Sachdev}.
When a controlling parameter changes across critical point, some
properties of the many-body system will change dramatically. Many
results show that entanglement existed naturally in the spin chain
when the temperature is at zero. The quantum entanglement of a
many-body system has been paid much attention since the entanglement
is considered as the heart in quantum information and
computation~\cite{Nielson,Bennett}. As the bipartite entanglement
measurement in a pure state, the von Neumann entropy~\cite{Bennett1}
in the antiferromagnetic anisotropic spin chain~\cite{Vidal} and
isotropic spin chain~\cite{Latorre} are investigated respectively.
By using the cross fields of quantum many-body theory and
quantum-information theory, von Neumann entropy is applied to detect
quantum critical behaviors~\cite{Preskill,Osborne,Amico,Gu,Kitaev}.
A typical example is that Osborne solved exactly one-dimensional
infinite-lattice transverse-field Ising model to obtain entropy by
the Jordan-Wigner transform. The entropy predicts the quantum phase
transition successfully~\cite{Osborne}. Moreover, another concept
from quantum information science, the ground state fidelity has been
used to qualify quantum phase transitions in the last few
years~\cite{Zhou01,Quan,Gu1,Abasto,Oelkers,Cozzini,Venuti,Buonsante,You,Yang,Zhou02,Tzeng,Ren}.
Because the fidelity is a measure of similarity between states, the
fidelity should drop abruptly at critical points as a consequence of
the dramatic changes in the structure of the ground states,
regardless of what type of internal order is present in quantum
many-body states. This result is the orthogonality of different
ground states due to state distinguishability. Many results shown
that the fidelity and the entanglement entropy have similar
predictive power for identifying quantum phase transitions in the
most systems~\cite{Tzeng,Ren}. However, the ground state fidelity is
a model-dependent indicator for quantum phase transitions. It cannot
be used to characterize the Berezinskii-Kosterlitz-Thouless (BKT)
transition~\cite{Chen,Chen01}. The BKT-like transition occurs at
Heisenebeg isotropic spin chain with next-nearest-neighbor and in
antiferromagnetic anisotropic Heisenberg model with
$\Delta=1$~\cite{Chen}. Entanglement, fidelity and their relations
with quantum phase transition in high spin chain, just like spin-1
chain need to be further investigated.

In this paper, the fidelity and entanglement in the easy-axis spin-1
chain are numerically investigated by using the density-matrix
renormalization-group (DMRG) technique. In section II, the
Hamiltonian of easy-axis spin-1 chain and its critical property are
presented. In section III, the effects of anisotropic interaction on
fidelity is investigated and its relation with quantum phase
transition is analyzed. The effects of anisotropic interaction on
entanglement entropy is calculated and its relation with quantum
phase transition is analyzed in section IV. At last, a discussion
concludes the paper.

\section{Hamiltonian and its critical property}

It is known that there is a gap in the spectrum of isotropic
Heisenberg spin-1 chain in the thermodynamic limit. The gap between
the ground state and the first excited state energy is usually
called Haldane gap~\cite{Haldane}. The Hamiltonian of an anisotropic
Heisenberg antiferromagnetic spin-1 chain of $N$ sites can be given
by

\begin{equation}
\label{eq1} H=J\sum_{i=1}^{N-1}(S^x_{i}S^x_{i+1}+S^y_{i}S^y_{i+1}+\Delta S^z_{i}S^z_{i+1}),\\
\end{equation}
where $S^{\alpha}_i(\alpha=x, y, z)$ are spin operators on the
$i$-th site and $N$ is the length of the spin chain. The parameter
$J>0$ denotes the antiferromagnetic coupling and $J=1$ is considered
in the paper. The parameter $\Delta$ is anisotropic interaction, $0
<\Delta < 1$ is the easy-plane anisotropy and $\Delta > 1$ is the
easy-axis anisotropy. It is predicted that a novel phase
$\Delta_{C1}\in[0\sim0.2]< \Delta< \Delta_{C2}=1.17$ appears between
the XY phase and the N\'{e}el phase~\cite{Botet,Nomura,Sakai}. Many
results suggested that a BKT transition occurs at $\Delta_{C1}$,
while that the transition at $\Delta_{C2}$ belongs to the 2D Ising
universality class~\cite{Affleck,H,Takahashi,White01}. Since the
fidelity cannot detect a BKT-like phase
transition~\cite{Chen,Chen01}, we only concentrate on quantum phase
transition which happens at easy-axis anisotropy. It is found that
the Hamiltonian in Eq. (1) can be transformed conveniently to give
\begin{equation}
\label{eq2} H=J\sum_{i=1}^{N-1}[p(S^x_{i}S^x_{i+1}+S^y_{i}S^y_{i+1})+S^z_{i}S^z_{i+1}],\\
\end{equation}
where $p=1/\Delta$. The critical point of transition from Haldane
spin liquid to N\'{e}el spin solid is around $p_c\simeq0.85$. It is
noted that the model of Eq. (2) is invariant for $p\rightarrow-p$,
so that the results in the paper can be equally applied to a system
with a ferromagnetic coupling on the $xy$ plane.

\section{Fidelity susceptibility}

The ground state fidelity can be applied to detect the existence of
the quantum phase transitions. A general Hamiltonian of quantum
many-body system can be written as $ H(\lambda)=H_0+\lambda H_I $
where $H_I$ is the driving Hamiltonian and $\lambda$ denotes its
strength. If $\rho(\lambda)$ represents a state of the system, the
ground state fidelity between $\rho(\lambda)$ and
$\rho(\lambda+\delta)$ can be defined as
\begin{equation}
\label{eq3}
F(\lambda,\delta)=Tr[\sqrt{\rho^{1/2}(\lambda)\rho(\lambda+\delta)\rho^{1/2}(\lambda)}].
\end{equation}
If the state can be written as $\rho=|\psi\rangle\langle \psi|$, the
Eq. (3) can be rewritten as $ F(\lambda,\delta)=|\langle
\psi(\lambda)|\psi(\lambda+\delta)\rangle|$. Because
$F(\lambda,\delta)$ reaches its maximum value $F_{max}=1$ for
$\delta=0$, on expanding the fidelity in powers of $\delta$, the
first derivative $\frac{\partial F(\lambda,\delta=0)}{\partial
\lambda}=0$. By using the property, the fidelity can written by
\begin{equation}
\label{eq4}
F(\lambda,\delta)\simeq1+\frac{\partial^2F(\lambda,\delta)}{2\partial\lambda^2}|_{\lambda=\lambda'}\delta^2.
\end{equation}
Therefore, the average fidelity susceptibility $S(\lambda,\delta)$
can be given by~\cite{Cozzini,Buonsante}
\begin{equation}
\label{eq5}S(\lambda,\delta)=\lim_{\delta\rightarrow
0}\frac{2[1-F(\lambda,\delta)]}{N\delta^2}.
\end{equation}

It is well known that it is hard to calculate the ground state
fidelity because of the lack of knowledge of the ground state
function. For models that are not exactly solvable, most of
researchers resort to exact diagonalization to obtain the ground
state for small size. This method cannot precisely quantify the
quantum phase transition because the size of the system is too
small. Recently, the method of density-matrix renormalization-group
(DMRG)~\cite{white,U} can be applied to obtain the ground state of
the model. Moreover, the technique of calculating the overlap of two
different ground states by DMRG has been used~\cite{Qin,McCulloch}.
For high precision, the method in~\cite{Tzeng1} is used to calculate
the ground state fidelity susceptibility. We calculate $N$ up to
$80$ when $\delta=0.001$~\cite{Tzeng}. The total number of density
matrix eigenstates held in system block is $m=70$ in the basis
truncation procedure. The Matlab codes of finite size density-matrix
renormalization-group have double precision. They are performed with
three sweep in private computer, and the truncation error is smaller
than $10^{-10}$.

The ground state fidelity susceptibility $S$ is plotted as a
function of anisotropic parameter $p$ for different sizes in Fig. 1
\emph{\textbf{in dimensionless units}}. It is shown that there is a
peak in $S$. The peak of $S$ increases when the size increases. The
location of the peak decreases slightly to small value of $p$ as $N$
increases. The ground state fidelity $F$ is plotted as a function of
$p$ for different sizes in the inset of Fig. 1 \textbf{\emph{in
dimensionless units}}. It is seen that there is a sharp valley
corresponding to the peak of $S$. The critical point of fidelity
susceptibility $S$ and fidelity $F$ is about $p_c=0.85$. As we
known, the fidelity measures the similarity between two states,
while quantum phase transitions are intuitively accompanied by an
abrupt change in the structure of the ground state wave-function.
This primary observation motivates researchers to use fidelity to
predict quantum phase transitions. Up to now, the ground state
fidelity and fidelity susceptibility have been applied in various
many-body systems to detect quantum phase transitions
successfully~\cite{Zhou01,Quan,Gu1,Abasto,Oelkers,Cozzini,Venuti,Buonsante,You,Yang,Zhou02,Tzeng,Ren}.
It confirms further that the Haldane spin liquid$-$N\'{e}el spin
solid transition occurs at the point in our system.

\section{Entanglement entropy}

For comparison, the ground state entanglement entropy is also used
to detect the quantum phase transition. For density matrix $\rho$ of
any pure state, the entanglement entropy $E_{AB}$ between subsystem
$A$ and $B$ can be defined as

\begin{equation}
\label{eq6}
E_{AB}=-Tr(\rho_{A}\log_2\rho_{A})=-Tr(\rho_{B}\log_2\rho_{B}),
\end{equation}
where $\rho_{A(B)}$is the reduced density matrix obtained from
$\rho$ by taking the partial trace over the state space of subsystem
$B(A)$.

In order to avoid boundary effects, the entanglement entropy of two
neighboring central sites of large system needs to be calculated by
using the method of DMRG. The entropy of two neighboring central
sites labeled $E_{N/2,N/2+1}$ is plotted as a function of anisotropy
$p$ with sizes of $N=40, 60, 80$ in Fig. 2 \emph{\textbf{in
dimensionless units}}. The entanglement entropy of the two
neighboring central sites increases monotonously when the anisotropy
$p$ increases. It is known that the peak or discontinuity in
entanglement entropy indicates the quantum phase transition. The
peak does not occur at the quantum critical point. It seems that
this is due to the monogamy property~\cite{Amico}. We also calculate
the entanglement entropy between the right-hand $N/2$ contiguous
qubits and the left-hand $N/2$ contiguous qubits. The result is
similar with that shown in Fig. 2. For avoiding repetition, the
result is not plotted again. This means that the entropy cannot be
used to predict quantum phase transition here. However, the critical
properties can be captured by the derivatives of the entropy as a
function of the anisotropy $p$~\cite{Amico,Osterloh,Liu}.

It is noted that the transition from Haldane spin liquid to N\'{e}el
spin solid is continuous and belongs to a second order quantum phase
transition. It has been shown that the density matrix may be
continuous but its first-order derivative is singular at a second
order quantum phase transition point~\cite{Wu}. Consequently, we
calculate the first-order derivative of entropy $dE_{N/2,N/2+1}/dp$
at the ground state with $\delta=0.001$. The first derivative of the
central two sites entanglement entropy $dE_{N/2,N/2+1}/dp$ is
plotted as a function of anisotropy $p$ for different sizes in Fig.
3. It is seen that there is a peak in $dE_{N/2,N/2+1}/dp$. The
location of the peak in $dE_{N/2,N/2+1}/dp$ moves slightly to small
value of $p$ as $N$ increases. The critical value is about
$p_c=0.85$. It represents indeed the quantum phase transition from
Haldane spin liquid to N\'{e}el spin solid.

\section{Discussion}

In the paper, the fidelity susceptibility and entropy in an
easy-axis antiferromagnetic Heisenberg spin-1 chain are studied. By
using the density-matrix renormalization-group for the model, the
effect of anisotropic interaction on fidelity susceptibility and
entropy of large size is presented. Their relations with quantum
phase transition are investigated. It is shown that the quantum
phase transition from Haldane spin liquid to N\'{e}el spin solid is
clearly marked by the peak (or valley) of the fidelity
susceptibility (or fidelity). However, the entanglement entropy
cannot have similar predictive power for revealing quantum phase
transition in the system due to the monogamy property, while the
first-order derivation of entropy can be used to successfully detect
the quantum phase transition.

\vskip 0.6 cm

{\textbf{Acknowledgments}}

It is a pleasure to thank Yinsheng Ling, Jianxing Fang, and Xiang
Hao for their many helpful discussions. The financial support from
the National Natural Science Foundation of China (Grant No.
10774108) is gratefully acknowledged.

\newpage

\begin{figure}
\includegraphics[scale=0.7]{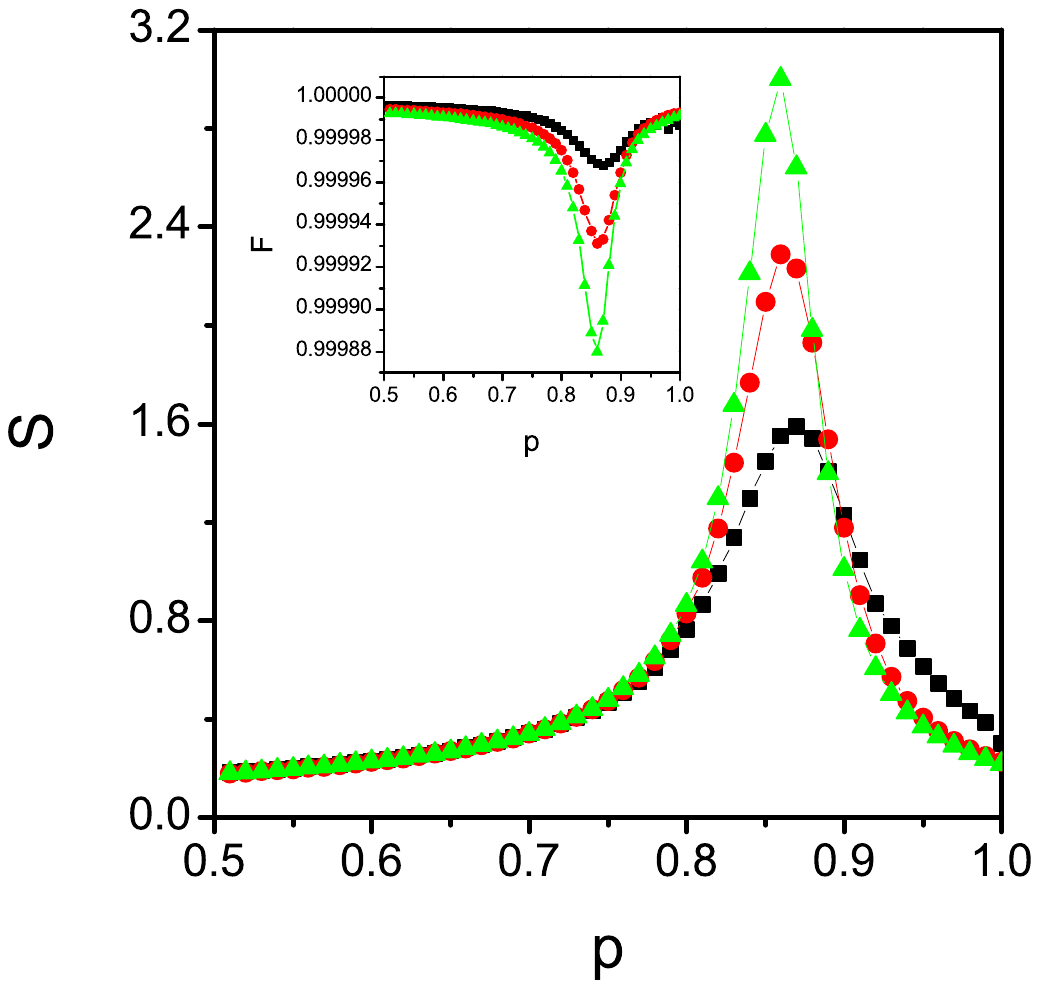}\caption{The fidelity susceptibility $S$
of the ground state is plotted as a function of anisotropy $p$ for
different size $N$. The fidelity $F$ is plotted in the inset. The
symbols are for
$N=40(\textcolor[rgb]{0.00,0.00,0.00}{\blacksquare}),
N=60(\textcolor[rgb]{0.98,0.00,0.00}{{\bullet}}),
N=80(\textcolor[rgb]{0.00,1.00,0.00}{{\blacktriangle}})$.}
\end{figure}

\newpage

\begin{figure}
\includegraphics[scale=0.7]{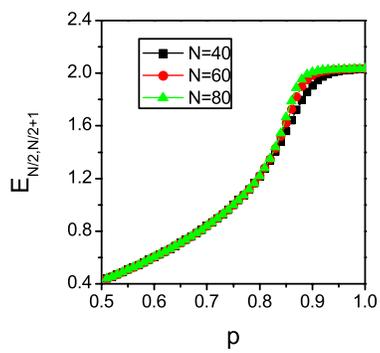}\caption{The two neighboring
central sites entanglement entropy $E_{N/2,N/2+1}$ is plotted as a
function of anisotropy $p$ for different size $N$.}
\end{figure}

\newpage

\begin{figure}
\includegraphics[scale=0.7]{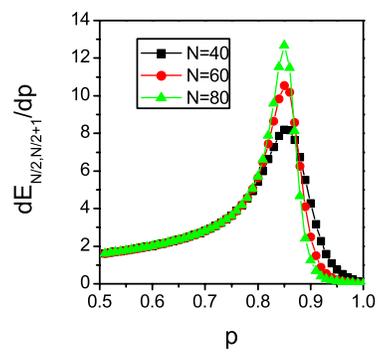}\caption{The first derivative of the two neighboring
central sites entanglement entropy $dE_{N/2,N/2+1}/dp$ is plotted as
a function of anisotropy $p$ for different size $N$.}
\end{figure}

\end{document}